# Refraction enhancement in plasmonics by the coherent control of plasmon resonances

Ali Panahpour[1], Abolfazl Mahmoodpoor[1], Andrei V. Lavrinenko[2,†]

[1]*Javan Laser Company Ltd., 1464764511, Tehran, Iran*
[2]*DTU Fotonik—Department of Photonics Engineering, Technical University of Denmark, Ørsteds pl. 343, DK-2800 Kongens Lyngby, Denmark*
[†]*Corresponding author: alav@fotonic.dtu.dk*

A plasmonic nanoantenna probed by a plane-polarized optical field in a medium with no gain materials can show zero absorption or even amplification, while exhibiting maximal polarizability. This occurs through coupling to an adjacent nanoantenna in a specially designed metamolecule, which is pumped by an orthogonal optical field with $\pi/2$ phase shift. The introduced scheme is a classical counterpart of an effect known in quantum optics as enhancement of the index of refraction (EIR). In contrary to electromagnetically induced transparency (EIT), where the medium is rendered highly dispersive at the point of zero susceptibility and minimum absorption, in the EIR the system exhibits large susceptibility and low dispersion at the point of zero or negative absorption. The plasmonic analogue of the EIR allows for coherent control over the polarizability and absorption of plasmonic nanoantennas, offering a novel approach to all optical switching and coherent control of transmission, diffraction and polarization conversion properties of plasmonic nanostructures, as well as propagation properties of surface plasmon polaritons on metasurfaces. It may also open up the way for lossless or amplifying propagation of optical waves in zero-index to high refractive index plasmonic metamaterials.

Optical response of surface plasmons are mostly governed by metal and ambient medium parameters, geometry of structures and also by plasmon hybridization, which can result in novel resonance line-shapes, enabling the plasmonic systems to mimic some quantum optical effects such as Fano interference and EIT [1-5]. The functionality of the plasmonic nanostructures is significantly improved through active plasmonics and specifically by all-optical and ultrafast control of the surface plasmon on the nanoscale, which has been realized by exploiting the phase and polarization degrees of freedom of the exciting waves [6-10].

All-optical coherent control of surface plasmons in single or coupled nanoantennas is of fundamental importance, since they can be viewed as building blocks of plasmonic metamaterials (MMs), metasurfaces and optical components such as sensors, switches, transistors and light sources. Coherent control over the linear [6,10], nonlinear [11] and spatially nonlocal [12] responses of plasmonic nanoantennas has been reported, as well as controlling the directionality of emission, scattering patterns and absorption in nanoparticles (NPs) involving nonlinear processes [13-15]. Moreover, coherent control of modal excitations in plasmonic metamolecules are investigated through adjusting the position of an exciting high-energy electron beam over a dolman style resonator [16] or by changing the relative phase of two orthogonally polarized light fields, exciting a triple nanorod structure [17]. Nevertheless, controlling a resonant plasmonic nanoantenna to exhibit strong polarizability and at the same time zero or negative absorption without involving materials with gain or nonlinear processes has not been reported so far.

Here, we propose a method for coherent control and manipulation of absorption and dispersion of a plasmonic nanoantenna, using only the phase and polarization degrees of freedom of light. The approach is based on a classical analogue of the quantum optics effect known as enhancement of the index of refraction, which occurs in atomic (or quantum dot) systems based on atomic coherence and quantum interference of different excitation pathways in variety of energy level schemes [18-21].

To see how the usual resonant dispersion is altered in EIR and how it differs from EIT, the corresponding dispersion curves are shown in Fig.1. The optical response of a two level atomic system, represented by the complex susceptibility $\chi$ is typically of the form shown in Fig.1a. Here, refraction (Re($\chi$)) is maximum at point *A*, where absorption (Im($\chi$)) is also very large. In contrast, the EIT, usually occurring in three level systems, modifies the dispersion profile to the typical form shown in Fig.1b. Here, absorption is minimal at point B corresponding to highest dispersion and zero susceptibility. While EIT is ideal for slow light applications and enhancement of nonlinear interactions, it is not so promising when strong electromagnetic response (large linear susceptibility) is required. Such a strong response accompanied with vanishing absorption arises in EIR, where the dispersion is modified to the form shown in Fig.1c.

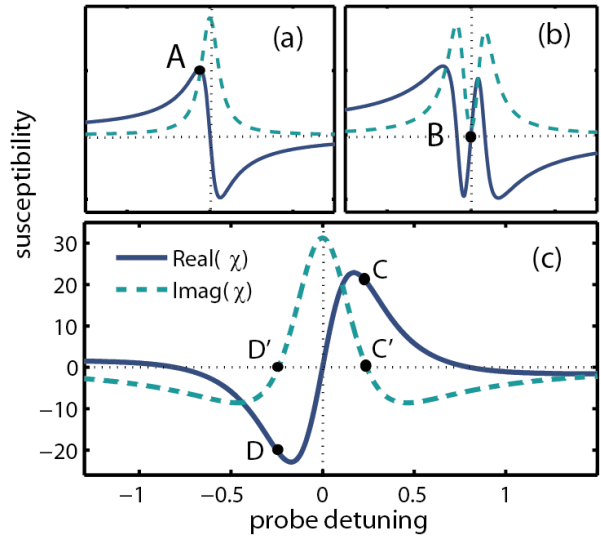

**FIG. 1:** Real and imaginary parts of susceptibility in a (a) two level system with standard dispersion curves, (b) coherent system showing EIT, (c) coherent system with enhanced refraction through EIR effect.

According to this figure, susceptibility (depicted as a function of the normalized probe detuning from the transition frequency of the excited level) is large at point C, where absorption totally vanishes (point C´). This results in huge lossless refractive indices in quantum coherent systems [21]. Additionally, susceptibility is negative at point D, where absorption is again zero (point D´). This spectral region is interesting for development of lossless negative or zero permittivity ($\text{Re}(\chi) = -1$) media, through proper tuning of the system parameters. Such implications are equally desirable in classical systems and can result in variety of novel applications in optics and nanophotonics, and specifically in plasmonics, where the conventional methods of loss compensation are still challenging [22-24].

Here, to introduce a simple toy model as a classical analogue of the EIR we study a mass-spring system consisting of two linearly coupled one-dimensional oscillators (Fig. 2) with coupling constant $C$ and damping factors $\gamma_1$ and $\gamma_2$.

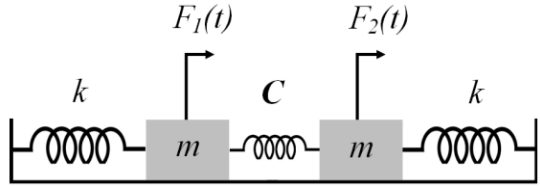

**FIG. 2:** Mass-spring analogue of the EIR consisting of two linearly coupled one-dimensional oscillators.

The complex amplitudes of the harmonic oscillators $x_j(t) = x_j \exp(-i\omega t)$ and driving forces $F_j(t) = F_j \exp(-i\omega t)$, are connected through:

$$\begin{pmatrix} x_1 \\ x_2 \end{pmatrix} = \begin{pmatrix} \delta_1 & -C \\ -C & \delta_2 \end{pmatrix}^{-1} \begin{pmatrix} F_1/m \\ F_2/m \end{pmatrix}, \qquad (1)$$

where, $\delta_j = \omega_0^2 - \omega^2 - i\gamma_j \omega$ and $j = 1, 2$. Assuming $F_1 = F_{01}$, $F_2 = \beta F_{02}$, $x_1 = x_{01}$ and $x_2 = \beta x_{02}$ where $F_{01}$ and $F_{02}$ are real numbers and $\beta = \exp(-i\theta)$ is a complex parameter depending on phase difference $\theta$ between the forces, we obtain (star sign denotes the complex conjugate):

$$x_{01} = \frac{1}{m} \cdot \frac{\delta_2 F_{01} + \beta C F_{02}}{\delta_1 \delta_2 - C^2}, \quad x_{02} = \frac{1}{m} \cdot \frac{\delta_1 F_{02} + \beta^* C F_{01}}{\delta_1 \delta_2 - C^2}, \quad (2)$$

and time averaged power absorbed by each oscillator is given by $W_j = \langle \text{Re}[F_j(t)] \times \text{Re}[\dot{x}_j(t)] \rangle_t = \frac{F_{0j}\omega}{2} \text{Im}(x_{0j})$.

The EIT effect can be reproduced in this system by adopting $\beta = 0$, which means that only the first oscillator is coupled to the external force and the second one is exclusively driven by virtue of coupling $C$. However, when both of the oscillators are driven by forces having some phase difference $\theta$, the system exhibits other interesting features. We assume parameters $F_{01}/m\omega_0^2 = 0.01$, $F_{02}/m\omega_0^2 = 1$, $C/\omega_0^2 = 0.06$, $\gamma_1 = \gamma_2 = 0.05\omega_0$, $\theta = \pi/2$ in (2) and plot the real and imaginary parts of the complex amplitudes $x_{01}$ and $x_{02}$ in Figs. 3a and 3b, respectively. By interchanging the values of the driving forces ($F_{01}/m\omega_0^2 = 1$ and $F_{02}/m\omega_0^2 = 0.01$) and other parameters as before, the complex amplitudes $x_{01}$ and $x_{02}$ are changed to the forms in Figs. 3c and 3d. Interestingly, the curves in Figs. 3a and 3d resemble the dispersion curves of a medium with enhanced index of refraction (Fig.1c). Similar results would be obtained if we used the phase difference $\theta = -\pi/2$. The only difference would be the interchange of curves in Figs. 3a and 3d.

The amplitude curves are not very sensitive to small deviations from $\theta = \pm\pi/2$. However, for large deviations of the phase difference from $\theta = \pm\pi/2$ especially getting closer to $\theta = 0$ or $\theta = \pm\pi$, the amplitude curves gradually convert to those of uncoupled oscillators.

According to Fig. 3 the complex amplitudes show negative absorption in specific spectral ranges around the resonance. However, using relations (2) it can be shown that the total power absorbed by both oscillators together $W = (\omega/2)[F_{01}\operatorname{Im}(x_{01}) + F_{02}\operatorname{Im}(x_{02})]$ is positive at all frequencies.

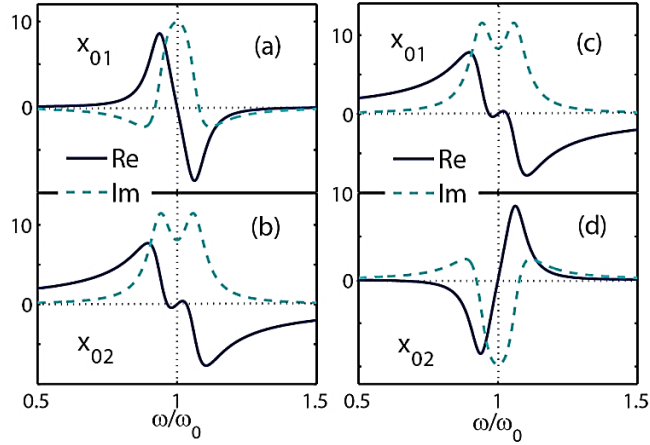

**FIG. 3:** Real and imaginary parts of $x_{01}$ and $x_{02}$ versus normalized frequency for $F_{01} \ll F_{02}$ (a), (b) and for $F_{01} \gg F_{02}$ (c), (d) correspondingly.

The complex amplitude curves in Figs. 3a and 3d inspire the application of microscopic coupled dipole oscillators with similar responses as the constitutive elements of a medium demonstrating the classical EIR. Plasmonic coupled nanoantennas as subwavelength constituent elements of the metamaterial can be an interesting platform, because the zero absorption feature in the EIR phenomenon can be exploited as a loss compensation mechanism in notoriously lossy plasmonic nanostructures.

To this end, we consider a specific metamolecule consisting of two coupled plasmonic nanoantennas which can be metal strips or nano-rods. To simplify analytical calculations, we consider two perpendicular prolate (cigar-shaped) spheroidal NPs shown in Fig. 4. They are excited by two plane waves with electric fields $E_1 = E_{01} e^{-i\omega t}$ and $E_2 = E_{02} e^{-i(\omega t + \theta)}$, having perpendicular polarizations in $x$ and $y$ directions, and frequency $\omega$ close to the localized surface plasmon resonance of the NPs along their longer semi-axes. The direction of propagation of the $x(y)$-polarized wave can be in any direction in a plane orthogonal to the $x(y)$ axis. For $E_{01} \ll E_{02}$ or $E_{01} \gg E_{02}$, and $\theta = \pi/2$, the curves for complex polarizability of the NPs are quite similar to plots in Fig. 3, corresponding to the conditions of $F_{01} \ll F_{02}$ and $F_{01} \gg F_{02}$. To study EIR in the condition of comparable pump and probe fields, we assume exciting fields with equal amplitudes. By taking $\theta = \pi/2$ the exciting fields with the same direction of propagation (Fig. 4) are actually the components of a circularly polarized wave.

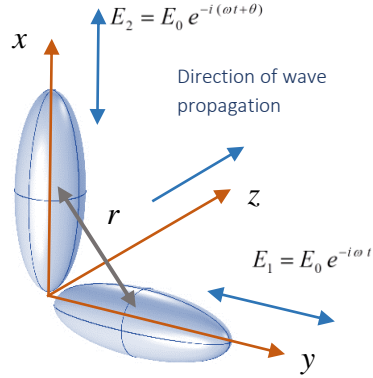

**FIG. 4:** A plasmonic molecule consisting of two perpendicular silver spheroidal nanoparticles excited by two orthogonal polarizations having a phase difference $\theta$.

The polarizability of a spheroidal NP of volume $v$ in the electrostatic limit is given by $\alpha = v(\varepsilon - \varepsilon_h)/[\varepsilon_h + L(\varepsilon - \varepsilon_h)]$ [25], where $\varepsilon$ and $\varepsilon_h$ are dielectric functions of the NP and host medium respectively, and $L = \frac{1-e^2}{e^2}\left(-1 + \frac{1}{2e}\ln\frac{1+e}{1-e}\right)$ is a geometrical factor in terms of eccentricity ($e^2 = 1 - b^2/a^2$) of the NPs [25]. We assume the nanoantennas to be silver NPs in vacuum ($\varepsilon_h = 1$) with the dielectric function of Johnson and Christy [26]. The size and separation of the NPs are assumed much smaller than the wavelength, so that they can be approximated as point dipoles with electrostatic approximation for their electric field.

The induced dipole moment along the longer semi-axis of each NP is affected by the superposition of the incident electric field and field induced by the dipole moment of the other NP. The projection of the latter on the longer semi-axis is proportional to dipole moment **p** with coefficient $C$ depending on the NPs separation and its orientation. Therefore, the polarizabilities of the coupled NPs (in terms of $\alpha$ and $\beta = e^{-i\theta}$) defined by the relations $p_1 = \varepsilon_0 \alpha'_1 E_0$ and $p_2 = \varepsilon_0 \alpha'_2 \beta E_0$ are obtained as $\alpha'_1 = \alpha \frac{1+\beta C \alpha}{1-C^2 \alpha^2}$ and $\alpha'_2 = \alpha \frac{1+\beta^* C \alpha}{1-C^2 \alpha^2}$.

Using the expression for the extinction cross section (ECS) of a dipole, excited by an incident plane wave $C_{ext.} = \left(k/\varepsilon_0 |E_0|^2\right) \text{Im}(\mathbf{E}^*_{inc} \cdot \mathbf{p})$ [27], the extinction of the linearly polarized components of the incident wave by the corresponding NPs can be calculated from $C_{ext.j} = k \, \text{Im}(\alpha'_j)$, ($j=1,2$). We consider spheroidal NPs with the semi-axis dimensions $b = c = 10 nm$ and $a = 30 nm$ in the configuration of Fig. 4. The real and imaginary components of the polarizabilities $\alpha'_1$ and $\alpha'_2$ (along the longer semi-axis of the NPs) as functions of wavelength for parameters $r = 50 nm$ and $\theta = \pi/2$ are plotted in Fig. 5.

We see that in the case of equal amplitude exciting fields, both of the NPs show negative absorption in different spectral ranges and exhibit zero absorption near the positive and negative peak values of their real polarizability. The individual and total ECSs of the NPs are represented in Fig. 5c. The ECS of each NP is negative in specific spectral ranges, but the total ECS is positive at all wavelengths as it is expected.

The polarizability values in Fig. 5a and 5b show that a 3D array of such metamolecules with the unit cell dimension of $100 nm$ (density of $10^{21}$ $m^{-3}$) can exhibit zero or negative permittivity at the lossless portion of the spectrum.

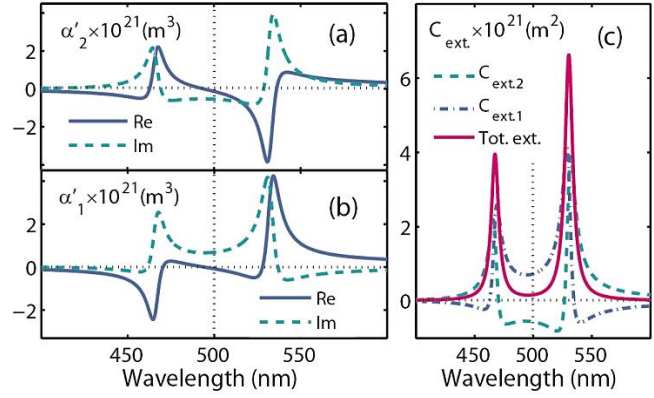

**FIG. 5:** Real and imaginary components of the polarizabilities (a) $\alpha'_2$ and (b) $\alpha'_1$ (of the coupled NPs in Fig. 4) as functions of wavelength, (c) The individual and total ECSs of the same NPs.

To support the analytical results, interaction of the incident circularly polarized wave with the NPs in Fig.4 is simulated in the frequency domain, based on the finite element method. The computational region is limited to the interior region of a perfect electric conductor (PEC) cylinder of diameter $D = 400 nm$ as a circular waveguide with input and output ports. The circularly polarized wave is approximated by a combination of two orthogonal $TE_{11}$ modes with a $\pi/2$ phase difference. The waveguide contains the coupled spheroidal NPs with the same geometrical dimensions as before, and with the dielectric function model of Johnson and Christy. The real and imaginary parts of the $x$ ($y$) component of the electric field near the NP oriented in $x$ ($y$) direction excited by the mode preceded (delayed) by $\theta = \pi/2$ in the waveguide is plotted in Fig. 6a (6b). The electric field is a sum of the contributions from the incident field and the dipole response of the coupled NPs. Therefore, the field near each NP can give insight into the effective polarizability of the particle.

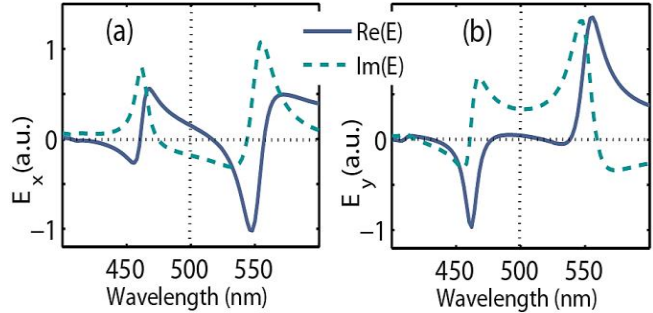

**FIG. 6:** Real and imaginary parts of (a) *x* and (b) *y* components of the electric field near the NPs oriented in the respective *x* and *y* directions, when the *x*-polarized field is preceded the *y*-polarized field by $\theta = \pi/2$ inside the waveguide.

The curves in Fig. 6 are quite similar to those in Figs. 5a and 5b, except for the spectral separation between the resonances, which are closer to each other in Fig. 5. The discrepancies between the simulation and analytical results can be attributed to the introduction of higher order (beyond the dipolar) interactions due to the close proximity of the NPs in simulation that also results in stronger coupling of the NPs compared to the conditions of analytical calculations where the NPs are approximated by point dipoles.

It can be inferred from the analytical and numerical results that the energy required for amplification of an incident linearly polarized wave by each NP is provided from the orthogonally polarized wave through the coupling mechanism with the other NP. This canalization of energy from one polarization to another through coupled NPs can be illustrated by tracing the streamlines of the power flow from the input to output ports of the circular waveguide. Fig. 7a (7b) shows the power flow streamlines around the NPs, corresponding to the wavelength of $\lambda = 500\,nm$ ($\lambda = 440\,nm$) at which the *x*-polarized (*y*-polarized) field is amplified, i.e. the input wave extinction due to interaction with the vertical (horizontal) NP is negative, while the extinction corresponding to the horizontal (vertical) NP is positive.

As expected, according to Fig. 7a, majority of the streamlines are headed for the horizontal absorptive NP, then redirected toward the vertical NP and finally steered toward the output port. Fig. 7c shows a closer view of the NPs, with a small number of streamlines to be traced with more clarity. Similarly, in Figs. 7b and 7d (corresponding to $\lambda = 440\,nm$) streamlines are mainly directed to the vertical NP with positive extinction and then redirected toward the output port by the horizontal NP having negative extinction. This property of power exchange between normal polarizations may find application in plasmonic polarization conversion metasurfaces [28, 29] with an augmented functionality of all optical control over the conversion efficiency.

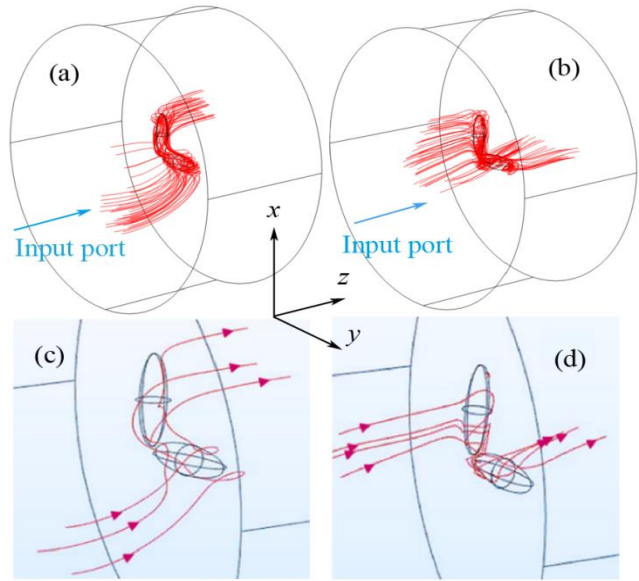

**FIG. 7:** Power flow streamlines around the NPs in the circular waveguide with the same parameters and geometrical dimensions as before and at the wavelengths (a, c) $\lambda = 500\,nm$ and (b, d) $\lambda = 440\,nm$.

Amplification of waves by the plasmonic coupled nanoantennas can be verified by the time-domain finite element method. Simulation of amplification in time domain is more challenging compared to the frequency domain simulation in the circular waveguide, because simultaneous absorption by one NP and amplification by another cannot be distinguished at the output port. Therefore, we use a configuration depicted in Fig. 8, where there are two separate waveguides with orthogonal electric fields corresponding to each NP.

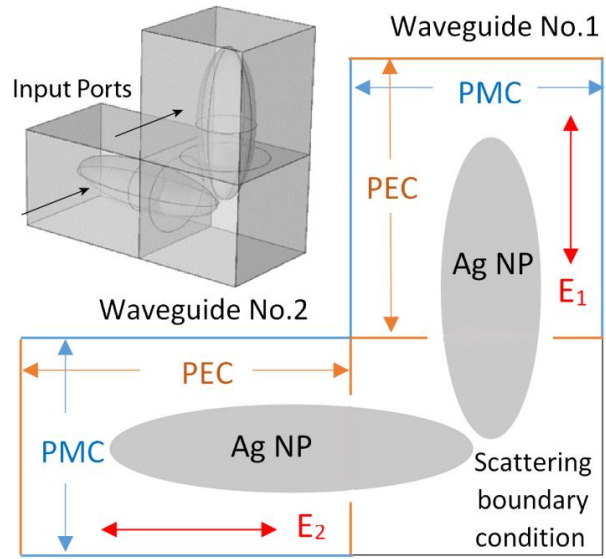

**FIG. 8:** Two waveguides with orthogonal electric fields corresponding to each NP of the metamolecule. The waveguides have two perfect electric and two perfect magnetic conductor side walls as well as input ports shown in the 3D picture with output ports in front of them.

The waveguides consist of two pairs of perfect electric and perfect magnetic conductor (PMC) sidewalls as well as input and output electromagnetic ports. The mutual coupling of the NPs is enabled by two small holes in the waveguides walls.

Electromagnetic fields with equal amplitudes and $\theta = \pi/2$ phase difference are fed into the waveguides from the input ports, and the field intensity is calculated as a function of time at the output ports. Field $E_1$ fed into waveguide No. 1 is preceding field $E_2$ in the other waveguide by $\theta = \pi/2$. Simulations with the NP parameters and dimensions as before ($\lambda = 500nm$) show (Fig.9) that the output field intensities deviated from those in the case $\theta = 0$ with nearly 10% growth for waveguide No.1 and nearly 10% attenuation for No.2, revealing the power transfer between waveguides.

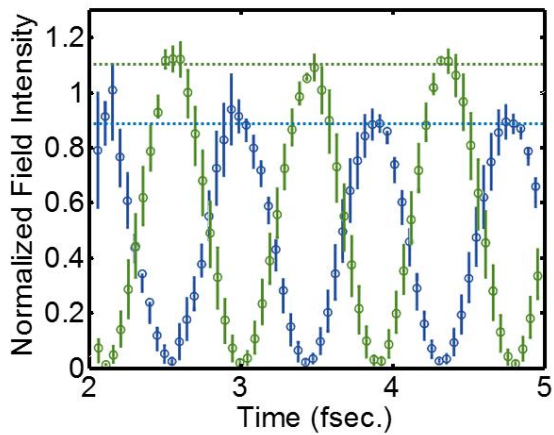

**FIG. 9:** Normalized field intensity at the output ports of the waveguides No. 1 (green curve) and No. 2 (blue curve) in Fig. 7.

The circles in Fig. 9, correspond to the calculations at different time steps and the bars represent the calculation errors due to the size of the mesh elements.

In summary, we introduced a classical mass-spring analogue and a plasmonic counterpart of the quantum optics effect of enhancement of index of refraction. The analogy shows that maximal polarizability of a plasmonic nanoantenna can be achieved accompanied by zero absorption or even amplification of the probe field. Therefore, it offers the way for development of lossless or amplifying plasmonic MMs and metasurfaces with tunable permittivities ranging from zero to maximal values. The classical EIR does not involve any gain or nonlinear materials, and based totally on polarizations and phases of electromagnetic waves. The concept can serve as an efficient loss compensation mechanism urgently required in plasmonics. Besides mitigating the loss problem, such scheme may find applications in all optical switching or optically tunable polarization conversion metasurfaces.

## Acknowledgements:

AVL acknowledges partial support from the Villum Fonden "DarkSILD" project (11116).

[4] N. Liu, L. Langguth, T. Weiss, J. Kästel, M. Fleischhauer, T. Pfau, and H. Giessen. "Plasmonic analogue of electromagnetically induced transparency at the Drude damping limit." *Nature Materials* 8, (2009): 758.

[5] P. Tassin, L. Zhang, R. Zhao, A. Jain, T. Koschny, and C. M. Soukoulis. "Electromagnetically induced transparency and absorption in metamaterials: the radiating two-oscillator model and its experimental confirmation." *Physical Review Letters* 109, (2012): 187401.

[6] M. Durach, A. Rusina, M. I. Stockman, and K. Nelson. "Toward full spatiotemporal control on the nanoscale." *Nano Letters* 7, (2007): 3145-3149.

[7] M. I. Stockman, "Ultrafast nanoplasmonics under coherent control." *New Journal of Physics* 10, (2008): 025031.

[8] M. Aeschlimann, M. Bauer, D. Bayer, T. Brixner, F. J. García De Abajo, W. Pfeiffer, M. Rohmer, C. Spindler, and F. Steeb. "Adaptive subwavelength control of nano-optical fields." *Nature* 446, 7133 (2007): 301.

[9] J.-S. Huang, D. V. Voronine, P. Tuchscherer, T. Brixner, and B. Hecht. "Deterministic spatiotemporal control of optical fields in nanoantennas and plasmonic circuits." *Physical Review B* 79, (2009): 195441.

[10] M. Sukharev, and T. Seideman. "Phase and polarization control as a route to plasmonic nanodevices." *Nano Letters* 6, (2006): 715-719.

[11] M. Abb, P. Albella, J. Aizpurua, and O. L. Muskens. "All-optical control of a single plasmonic nanoantenna–ITO hybrid." *Nano Letters* 11, (2011): 2457-2463.

[12] D. McArthur, B. Hourahine, and F. Papoff. "Coherent control of plasmons in nanoparticles with nonlocal response." *Optics Communications* 382 (2017): 258-265.